\documentclass[%
 aip,
 amsmath,amssymb,
 reprint,%
]{revtex4-1}
\usepackage{graphicx}
\usepackage{dcolumn}
\usepackage{bm}
\usepackage[mathlines]{lineno}
\usepackage[utf8]{inputenc}
\usepackage[T1]{fontenc}
\usepackage{mathptmx}
\usepackage{float}
\usepackage[normalem]{ulem}
\usepackage{nicefrac}
\usepackage{xcolor} 
\usepackage{todonotes}
\usepackage{csquotes}
\usepackage{upgreek}
\usepackage{soul}
\usepackage{hyperref}

\definecolor{darkred}{rgb}{0.65, 0, 0}

\begin{document}

\preprint{AIP/123-QED}

\title{Piezoelectric-driven uniaxial pressure cell for muon spin relaxation and neutron scattering experiments}

\author{Shreenanda Ghosh}
\affiliation{Institute for Solid State and Materials Physics, Technical University of Dresden, \\D-01069, Germany}
\author{Felix Br\"uckner}
\affiliation{Institute for Solid State and Materials Physics, Technical University of Dresden, \\D-01069, Germany}
\author{Artem Nikitin}
\affiliation{Laboratory for Muon Spin Spectroscopy, Paul Scherrer Institute, CH-5232 Villigen, Switzerland}
\author{Vadim Grinenko}
\affiliation{Institute for Solid State and Materials Physics, Technical University of Dresden, \\D-01069, Germany}
\affiliation{IFW, D-01069 Dresden, Germany}
\author{Matthias Elender}
\affiliation{Laboratory for Muon Spin Spectroscopy, Paul Scherrer Institute, CH-5232 Villigen, Switzerland}
\author{Andrew P. Mackenzie}
\affiliation{Max Planck Institute for Chemical Physics of Solids, D-01187 Dresden, Germany}
\author{Hubertus Luetkens}
\affiliation{Laboratory for Muon Spin Spectroscopy, Paul Scherrer Institute, CH-5232 Villigen, Switzerland}
\author{Hans-Henning Klauss}%
\email{henning.klauss@tu-dresden.de}
\affiliation{Institute for Solid State and Materials Physics, Technical University of Dresden, \\D-01069, Germany}
\author{Clifford W. Hicks}
\email{C.Hicks.1@bham.ac.uk}
\affiliation{Max Planck Institute for Chemical Physics of Solids, D-01187 Dresden, Germany}
\affiliation{School of Physics and Astronomy, University of Birmingham, Birmingham B15 2TT, United Kingdom}

\date{\today}
\begin{abstract}
We present a piezoelectric-driven uniaxial pressure cell that is optimized for muon spin relaxation and neutron scattering experiments, and that is operable over a wide temperature range including cryogenic temperatures. To accommodate the large samples required for these measurement techniques, the cell is designed to generate forces up to $\sim$1000 N, and to minimize the background signal the space around the sample is kept as open as possible. We demonstrate here that by mounting plate-like samples with epoxy, a uniaxial stress exceeding 1 GPa can be achieved in an active volume of 5 mm$^{3}$. We show that for practical operation it is important to monitor both the force and displacement applied to the sample. Also, because time is critical during facility experiments, samples are mounted in detachable holders that can be rapidly exchanged. The piezoelectric actuators are likewise contained in an exchangeable cartridge.
\end{abstract}

\maketitle
\section{Introduction}\label{sec:intro}

Uniaxial stress is an established method for probing  materials. It can induce tunable changes to electronic structure without introducing disorder. Many probes of condensed matter
physics can be combined with uniaxial stress in a technically straightforward way\cite{ Meingast91, Taniguchi13, Tamatsukuri16}: in contrast to hydrostatic stress, there is no pressure medium, and so the sample can remain exposed. Here, we discuss an apparatus for muon spin relaxation ($\mu$SR) and neutron scattering. These methods typically require larger samples than other probes, and therefore higher force to achieve a specified stress.
We show here that it is practical to use a piezoelectric-based apparatus to apply tunable GPa-scale uniaxial stresses to large samples. 

Uniaxial stress was traditionally applied by compressing samples between two anvils \cite{Takeshita04}. However, GPa-level uniaxial stresses with high stress homogeneity can be achieved more reliably by preparing samples as long, narrow beams and embedding their ends in epoxy~\cite{Hicks14RSI, Barber19RSI}. Force is then transferred to the sample through shear stress applied through the epoxy to the sides of the sample,
rather than through compressive stress applied to the sample ends. The epoxy constitutes a conformal layer that eliminates points of high stress concentration. Furthermore, the low elastic moduli of epoxies relative to typical samples mean that force is applied to the sample over a relatively large area, reducing the maximum shear stress at the sample-epoxy interface.

In this way, uniaxial stresses of up to $\sim$2~GPa have been applied successfully to samples of thickness up to $\sim$0.1~mm \cite{Barber19RSI}. Although increasing the size of the sample places greater stress on the epoxy, we show here that by using this epoxy-based sample mounting a uniaxial stress of at least 1 GPa is nevertheless achievable in samples large enough for practical $\mu$SR and neutron scattering experiments. Achieving this stress requires careful sample preparation that must in practice be done well in advance of beamtimes. Therefore, a major practical consideration in making such measurements possible is the
design of a suitable sample holder that safely transmits force to the sample. This holder must be attachable to a device that generates large forces, and the attachment process must not damage the sample. Exchanging holders must also be a
rapid process, to minimize loss of time during a beamtime. \par There are a number of possibilities to generate the force. To apply force at cryogenic temperature, helium-filled bellows can be
used~\cite{Fobes17}, or transfer rods from room temperature. The approach we take here is to use piezoelectric actuators: along with a suitable sample holder, we present a design for a
piezoelectric-based force generator of diameter 23~mm that generates forces of up to $\sim$1000~N. A subset of the present authors presented an earlier version of this system in Ref.\citenum{Hicks18}.
Since then, a number of practical improvements have been made, along with demonstration of the capabilities of this cell.

\begin{figure*}
\includegraphics[width=180mm,scale=1.5]{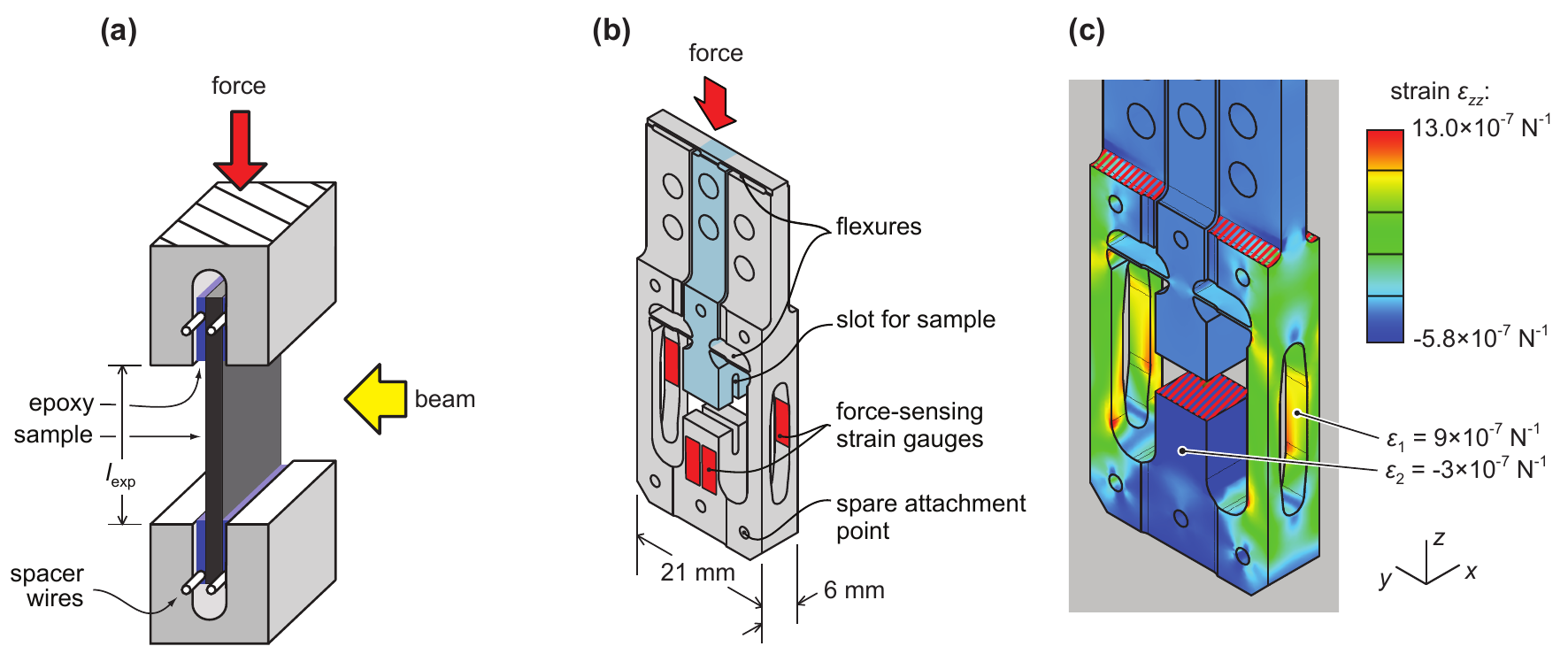}
\caption{\label{fig:holder}\textbf{(a)} The essential experimental setup. Force is applied vertically, along the length of a plate-like sample. Force is transferred into the sample through epoxy
layers. Spacer wires may be inserted to control the thickness of the epoxy as it cures. \textbf{(b)} Schematic of the sample holder.  This design offers access to the sample from all four sides. \textbf{(c)} Finite
element analysis of the $z$-axis strain $\varepsilon_{zz}$ within the sample holder under an applied load. In this simulation, the load is applied at the red hatched surfaces. The strain at the mounting locations of the strain gauges, per newton of applied force, is indicated.}
\end{figure*}

\section{The sample holder}
\label{sec:Holder} 

In a $\mu$SR experiment, spin-polarized muons are implanted into the sample, where each of them interacts with its local magnetic field. By collecting statistics on the positron emission direction as
the muons decay, the evolution of the average polarization with time can be determined, providing information on the internal magnetic fields of the sample. $\mu$SR measurements have been performed on
hydrostatically pressurized samples by using muons sufficiently energetic (typically $\sim$50~MeV) to penetrate the walls of the pressure cell\cite{{Khasanov16}}. In our uniaxial pressure cell the
sample is directly exposed to the beam, and instead so-called ``surface muons'' can be used. These have a narrower energy distribution, centred around 4.1~MeV, and so a better-controlled stopping
distance, of typically 0.1--0.3~mm~\cite{PSI, Sonier00}. (These muons are termed ``surface muon'' because they originate from decay of pions that come to rest near the surface of the graphite target
in which the pions are created.)

The essential sample configuration is illustrated in Fig.~\ref{fig:holder}(a). The sample is plate-like. For $\mu$SR measurements, the key advantage of a plate-like geometry is that it maximizes the area facing the muon
beam for a given sample volume, maximizing the number of muons that are intercepted. For neutron scattering the shape is less important because neutrons penetrate more deeply, however a plate-like
geometry is still useful as it maximizes the area of the epoxy layers, through which force is applied to the sample. For compressive loads, the upper limit on the sample length-to-thickness ratio is
set by the buckling instability. Euler's formula for the buckling load $F$ on a plate-like sample of length $l$, width $w$, and thickness $t$ is
\[
F = \frac{\pi^2 E t^3 w}{3 l^2},
\]
where $E$ is the Young's modulus of the sample. Note that because the epoxy layers are deformable, $l$ in this expression will be somewhat longer than the actual exposed sample length, $l_\text{exp}$
in Fig.~\ref{fig:holder}(a).

Our holder design is shown in Fig.~\ref{fig:holder}(b). The sample is mounted between an outer portion of the holder (colored gray in the figure) that may be considered fixed, and an inner portion (colored blue)
that can move. This moving portion is guided by four flexures. These flexures together present a high spring constant against all motions except the desired longitudinal motion, and so provide
mechanical protection to the sample during handling. 

The upper portion of the holder slots between flanges of the force generator. The flanges are smooth and so the mating is frictional, which is suitable for use with a piezoelectric-based generator:
piezoelectric actuators can generate high force but only small displacements, and so it is desirable that generated displacements are transferred to the sample as efficiently as possible, without loss
due to mechanical play. To avoid inadvertently applying force to the sample during insertion, the holder can be inserted into the flanges of the generator from the side.
The dimensions of the slots that accept the sample can be adjusted to match the sample dimensions. There are windows in the sides of the holder, to facilitate fabrication of these slots through
electrostatic discharge machining. These windows are also a practical necessity for sample mounting: to achieve high forces, the thickness of the epoxy layers must be controlled, and one wants to be
able to inspect them after the epoxy is cured. 
\par The space around the sample is otherwise kept as open as possible, in order to minimize the count rate from material other than the sample. The portions of the holder that intrude into the beam can be
masked, with hematite or silver sheets for $\mu$SR, or cadmium sheets for neutron scattering. 
\begin{figure*}
\includegraphics[width=180mm,scale=1.5]{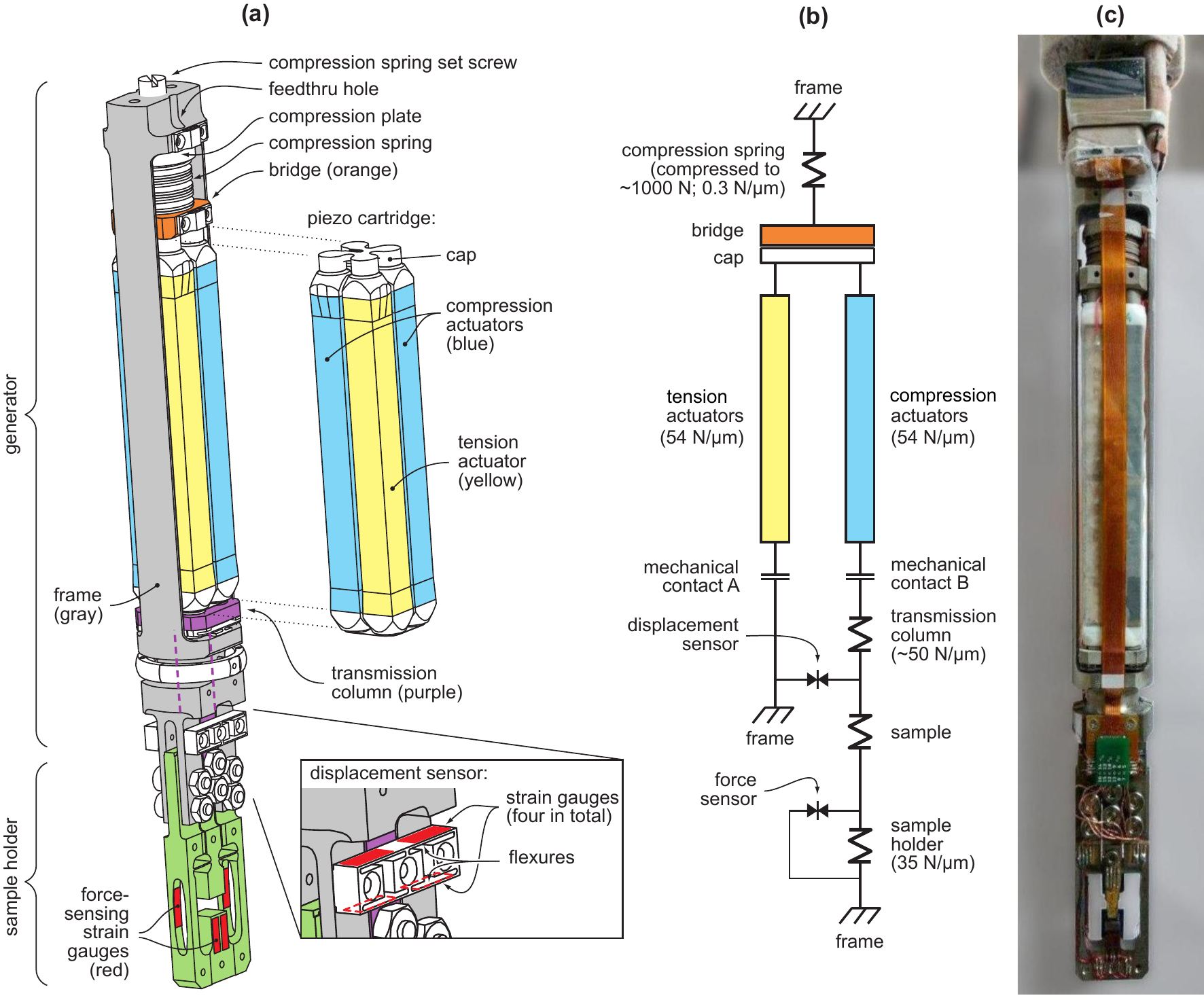}
\caption{\label{fig:device}\textbf{(a)} Schematic of the uniaxial pressure cell, comprising a generator and a sample holder. The gray- and green-coloured parts are respectively the Ti frame of the
generator and the detachable sample holder. Force is transmitted from the actuators to the sample holder through a transmission column, colored purple in the figure. The bottom ends of the tension
actuators are anchored to the frame, through holes that pass through the top plate of the transmission column, while the bottom ends of the compression actuators are anchored to the transmission column. The actuators are contained within a removable cartridge. \textbf{(b)} Diagram of the mechanics of the cell. Each component is labelled by its approximate spring constant. The relative lengths of the tension and compression actuators then determine the fraction of this preload that is transferred to the sample. \textbf{(c)} Photograph of the cell.}
\end{figure*}
\par The struts on the side of the holder provide mounting surfaces for strain gauges, as illustrated in Fig.~\ref{fig:holder}(b). By measuring the elongation of the struts when a load is applied, the gauges
serve as a sensor of the force applied to the sample. As illustrated in Fig.~\ref{fig:holder}(c), the strain is largest on the inner surfaces of the struts, so two of the strain gauges are mounted to these surfaces. The gauges
are connected in Wheatstone bridge configuration. To complete the bridge, two further gauges are mounted to a portion of the holder that comes under compressive strain when the sample is compressed.

\section{The generator}
We now describe a generator that uses piezoelectric actuators to generate a compressive force of up to $\sim$1000~N. A schematic of this generator along with the sample holder is shown in Fig.~\ref{fig:device}(a).
It has a diameter of 23~mm and a length, including the sample holder, of $202$~mm. It was designed to fit into a $^3$He cryostat of the Dolly spectrometer installed at the $\pi$E1 muon beamline at the Swiss Muon Source.
\par The core mechanism of the generator is a sustained preload force of $\sim$1000 N, generated by a compression spring, applied to two sets of piezoelectric actuators. The relative length of the actuators determines the fraction of the preload that is transferred to the sample. The preload spring consists of Belleville springs, and is located at the top of the generator. The force is applied through a bridge plate to four piezoelectric actuators that are joined electrically into two pairs, the `compression' and `tension' actuators. The other end of the tension actuators is coupled to the frame of the generator, while the other end of the compression actuators is coupled, through a transmission column, to the sample. When the compression actuators are lengthened relative to the tension actuators, they carry a greater fraction of the preload, which in turn is transmitted to the sample. When the temperature of the apparatus is changed, the lengths of the compression and tension actuators change together, so there is minimal change in the load on the sample. A schematic diagram of the mechanical connections in the stress cell is shown in Fig.~\ref{fig:device}(b).

The preload is held by a set screw, that fixes the position of a compression plate that pushes on the preload spring. It is not practical to turn this screw against the preload, and so the top of the
cell incorporates tapped holes to which an external press can be attached, and feed-thru holes through which the press can depress the compression plate. 

The actuators are contained in a cartridge that can be exchanged when the preload is released. A consequence of the compactness of this cell is that the actuators must be driven to high voltages. The actuators can in practice be driven to voltages well above the manufacturer's specified maximum, especially at low temperatures, and the overall design decision we make here is to accept the increased risk of actuator breakdown or fracture in exchange for a more compact cell. It is therefore desirable that the actuators be exchangeable during a beamtime, should one or more fail. In the cartridge, the actuators are joined at their top ends by a stiff cap plate, and at their bottom ends by a flexible foil that allows the tension and compression actuators to have different lengths. This cartridge is one of the key improvements on the cell presented in Ref.\citenum{Hicks18}.

The actuator cartridge is held in place by the preload force, without any adhesive. The interfaces at the bottom end of the cartridge therefore constitute contact interfaces that open when the load
becomes tensile. This prevents any individual actuator from coming under a tensile load, which they are not designed to withstand~\cite{PIinfo}. The interfaces also serve as limiters on the force that the generator can apply. When the load on the compression actuators falls to zero, the interface under the
compression actuators [labelled `mechanical contact B' in Fig.~\ref{fig:device}(b)] opens, such that the load on the sample cannot become tensile. Conversely, when the load on the sample reaches the full preload, the interface under the tension actuators opens (`mechanical contact A'), and the force on the sample goes no higher. The ability to bring the load on the sample reliably to zero is highly useful for control measurements. Although the force-sensing strain gauges on the sample holder can be calibrated in advance, strain gauges are prone to drift and are affected by temperature, introducing uncertainty.

The accuracy of this zero-force calibration is limited by the spring constants of the flexures. The spring constant of the flexures in the holder is 0.14~N/$\mu$m, and of those in the generator 0.18~N/$\mu$m. (Both values are obtained from simulation.) If the sample contracts thermally differently from the titanium of the holder, it will apply displacement to these flexures, resulting in a force on the sample. A 10~$\mu$m thermal displacement, for example, results in a $\approx 3$~N force on the sample.

The generator also incorporates a set of four strain gauges, connected in a Wheatstone bridge configuration, to measure the displacement between the frame of the generator and the transmission column. They are attached to flexures that guide the motion of the transmission column. More details are given in the \hyperref[sec:Appendix]{Appendix}. While the force sensor on the sample holder is intended to be the primary sensor of the state of the sample, the displacement sensor
provides essential information on the mechanical state of the sample. For example, a fracture appears as an increase in displacement accompanied by a decrease in force.

The frame of the generator is fabricated from commercial pure titanium. Flexures are fabricated from the alloy Ti$_{0.90}$V$_{0.04}$Al$_{0.06}$, which has a higher elastic limit. We avoid extensive use of this alloy because it has a higher superconducting critical temperature (3.5~K) and much worse thermal conductivity than commercial pure titanium, which itself is not a good
conductor. To reliably reach temperatures below 1~K, it was necessary to add copper foils joining the cold plate of the cryostat and the sample holder; with this provision a base temperature of 0.4~K was reached.

\section{Analysis of the generator}
\begin{figure}
\includegraphics[width=\columnwidth]{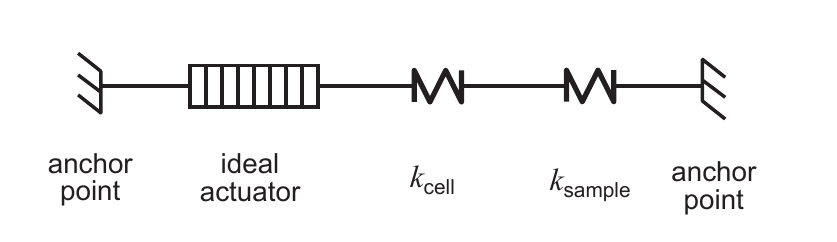}
\caption{Basic model of a piezoelectric-driven uniaxial pressure cell.}
\label{fig:overview_schematic}
\end{figure}

The performance of a piezoelectric-driven stress apparatus can be characterized by two quantities, the zero-load maximum displacement and the zero-displacement maximum force. The former is the displacement that can be applied to a sample of zero spring constant, and the latter is the force that can be applied to a sample of infinite spring constant. A conceptual diagram is illustrated in
Fig.~\ref{fig:overview_schematic}. The pressure cell can be modelled as an ideal actuator, that generates a specified displacement regardless of the resisting force, in series with a spring, which represents the internal spring constant of the cell. It is analogous to the voltage source and internal resistance model of batteries. 

The actuators each have dimensions $7 \times 7 \times 72$~mm\cite{PI}, and for the actuator material we take a Young's modulus of 40~GPa, determined from their room-temperature specifications~\cite{PIinfo}.
These parameters yield a spring constant of $\sim$27~N/$\mu$m for each individual actuator. The spring constants of other elements of the cell were estimated through finite element analysis. Combining all these elements, the spring constant of the cell can be estimated:
\begin{align*}
k_\text{cell}{^{-1}} &= (k_\text{act} + k_\text{spring}){^{-1}} + k_\text{act}{^{-1}} + k_\text{trans}{^{-1}}+k_\text{holder}{^{-1}}, 
\end{align*}
where $k_\text{act} = 54$~N/$\mu$m is the spring constant of each pair of actuators, $k_\text{trans}=50$~N/$\mu$m is that of the transmission column, $k_\text{holder}=35$~N/$\mu$m is that of the holder, and $k_\text{spring}=0.3$~N/$\mu$m is the spring constant of the preload spring. This expression gives us $k_\text{cell}\approx 12~\text{N/$\mu$m}$. \par Data reported in Ref.~\citenum{Barber19RSI} show that at 1.5~K and $-300$~V the strain in the actuators is $\sim -7 \cdot 10^{-4}$, and at $+400$~V, $+8 \cdot 10^{-4}$.  Assuming therefore maximum voltages of
$-300$~V on the tension actuators and $+400$~V on the compression actuators, a maximum displacement of 108~$\mu$m is generated. This is the zero-load maximum displacement, and it is a soft maximum because the voltages can be further increased, with gradually increasing risk of fracture in the actuators.  The zero-displacement maximum force is $\sim 108$~$\mu$m $\times$ 12~N/$\mu$m = 1300~N. Because this maximum exceeds the preload, in reality mechanical contact A would open before this force is reached. 

The preload force increases as the cell is cooled. With cooling to cryogenic temperatures, the actuators lengthen by $\approx$0.1\%, and the frame of the generator contracts by 0.15\%, giving a thermal displacement of $\approx$ 180 $\mu$m. The spring constant of the preload spring is selected so that under this additional compression the preload does not increase too much; here, the increase is $\approx$ 60 N. In addition, the elastic moduli of the material of the preload spring (beryllium copper) will increase with cooling, further increasing the preload.
\section{Calibration of the force and displacement sensors}
The force sensor conversion constant $\alpha_F$, in units of $\Omega$/N, can be estimated through finite element analysis. $\alpha_F = G R(\varepsilon_1/2 - \varepsilon_2/2)$, where $G\approx 2$ is the gauge factor of the gauges, $R = 350$~$\Omega$ is the gauge resistance\cite{Gauge}, and $\varepsilon_1$ and $\varepsilon_2$ are the
strains at the two locations of the strain gauges, indicated in Fig.~\ref{fig:holder}. From finite element analysis, $\varepsilon_1 = 9 \cdot 10^{-7}$~N$^{-1}$ and $\varepsilon_2 = -3 \cdot 10^{-7}$~N$^{-1}$, yielding $\alpha_F = 4.2 \cdot
10^{-4}$~$\Omega$/N. The calibration can also be determined experimentally by hanging known weight from the holder. Doing so, we obtained at room temperature $\alpha_F = 4.6 \cdot
10^{-4}$~$\Omega$/N. The Young's modulus of titanium increases by $14\pm 1$\% between 300~K and $T \rightarrow 0$~\cite{Ekin06}, and so, neglecting changes in $G$, $\alpha_F$ will decrease to $4.0 \cdot 10^{-4}$~$\Omega$/N at cryogenic temperature. Force values reported in this paper are obtained using this value of $\alpha_F$. The stress in the sample is then the force divided by sample cross-sectional area.
The displacement sensor can be calibrated using an interferometer or a microscope to measure the applied displacement. Here, a conversion constant of $\alpha_D = 5.7 \cdot 10^{-3}$~$\Omega$/$\mu$m
was obtained at room temperature by using an interferometer to measure the displacement in an empty sample holder. $\alpha_D$ is not affected by the elastic moduli of the flexures to which the displacement-sensing gauges are attached, so $\alpha_D$ is not expected to change strongly with temperature. The setups used for force and displacement sensor calibration are shown in the \hyperref[sec:Appendix]{Appendix}.
\begin{figure}
\includegraphics[width=\columnwidth]{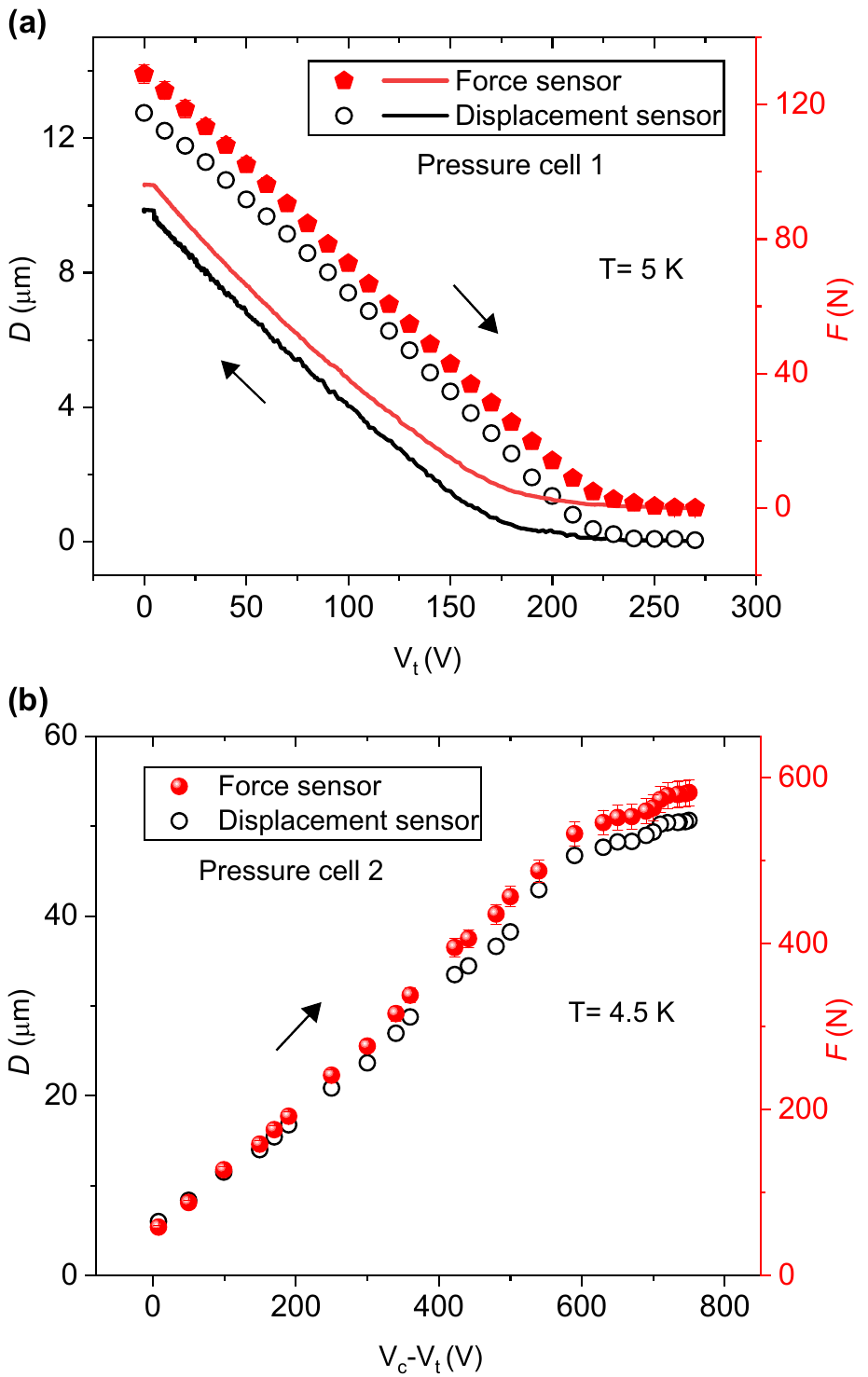}
\caption{Force limits of the cell. \textbf{(a)} The lower limit. D is the displacement measured by the displacement sensor, and F the force applied to the sample. The tension actuator voltage $V_t$ was varied while the compression actuator voltage $V_c$ was held at 0~V; due to hysteresis in the actuators, the increasing-$V_t$ and decreasing-$V_t$ curves do not overlap. When the load on the sample is reduced to zero, mechanical contact B (see Fig.~\ref{fig:device}) opens, and the force and displacement sensor readings stop changing. This flattening can therefore be identified as the zero-force point. 
\textbf{(b)} The upper limit. When the load on the sample reaches the preload force, mechanical contact A opens, and the load on the sample does not increase further. Here the preload springs were compressed only to $\sim$600~N rather than the maximum of $\sim$1000~N.}
\label{fig:cell_limits}
\end{figure}
\par Although the displacement sensor is intended primarily to monitor for damage to the sample and/or epoxy as the applied stress is increased, it can also be used for low-precision estimates of the elastic sample strain. If deformation of the epoxy and of the portions of the sample embedded in the epoxy is neglected, then the strain in the sample can be taken as $D_\text{sam}/l_\text{exp}$, where $D_\text{sam}$ is the displacement applied to the sample. In practice, epoxy deformation is usually not negligible, and this quantity is an upper bound on the strain in the sample. $D_\text{sam}$ is related to the displacement $D$ measured at the displacement sensor by

\[D_\text{sam} = D - F/k_\text{disp},\]where $k_{disp}$ is the spring constant of all mechanical elements between the anchor points of the displacement sensor and the sample. We used the
interferometer again, on a holder that had a sample mounted, to obtain $D_\text{sam}$. $F$ and $D$ were determined simultaneously from the displacement and force sensors, using the above-mentioned
conversion constants. Combining, we obtained $k_\text{disp} = 27$~N/$\mu$m, which is, as expected, slightly lower than the spring constant of the sample holder alone, 35~N/$\mu$m.

\section{Demonstrations of the cell}

Most of the demonstrations reported here were done on the unconventional superconductor Sr$_2$RuO$_4$\cite{Mackenzie17, kivelson2020proposal}. Under a uniaxial stress of $\sigma = -0.71 \pm 0.08$~GPa
(where negative values denote compression) along a $\langle 100 \rangle$ lattice direction at 5 K, Sr$_2$RuO$_4$ is driven through a transition in its Fermi surface topology, at which its
low-temperature electronic properties change drastically \cite{Barber2019PRB}. Its superconducting critical temperature $T_c$ rises from 1.5~K at $\sigma = 0$ to a peak of 3.5~K at $\sigma = -0.75 \pm
0.08$~GPa \cite{Barber2019PRB}.  All of the data reported here were recorded over the course of $\mu$SR measurements on the $\pi$E1 beamline of the Paul Scherrer Institute. We first demonstrate the
mechanical limits of the generator. In all tests we report here, the samples were inserted into the cell and cooled with $(V_c, V_t) = (0,0)$~V, where $V_c$ and $V_t$ are respectively the voltages on
the compression and tension actuators. We generally used Stycast\textsuperscript{\textregistered} 2850 to mount samples, including the sample in Ref.\citenum{Grinenko19} in which a stress exceeding 1
GPa was reached, setting the epoxy layer thickness to 60–100~$\mu$m. With this epoxy thickness, according to the model of Ref.\citenum{Hicks14RSI} strain is transferred to a 0.4 mm-thick sample of
Sr$_2$RuO$_4$ (which has a Young's modulus of 160 GPa~\cite{Barber2019PRB}) over a $\sim$700 $\mu$m length scale. We generally aim to have a length of sample at least 2--3 as large as this embedded in
the epoxy at each end. So, for example, to obtain a 5-mm length of sample directly exposed to the beam, the total sample length in this example would be about 9~mm: the 5~mm exposed portion in the
middle, and then about 2~mm on each end embedded in the epoxy.  Some samples were also mounted with Masterbond\textsuperscript{\textregistered}  29LPSP, an epoxy that wicks more effectively into the
gap between the sample and holder than Stycast; a similar layer thickness was used. So far, we have not rigorously compared the low-temperature stress limits of different epoxies. 

In Fig.~\ref{fig:cell_limits}(a), a determination of the zero-force point reading of the $F$ and $D$ sensors is shown. $V_c$ was held at zero, and $V_t$ increased until the readings from these sensors
stopped changing. This is the point at which the mechanical contact under the compression actuators opened, and the load on the sample was reduced to zero. In panel (b) the opposite limit is
demonstrated: when the compression actuators are extended sufficiently, the preload force is transferred fully to the sample, and the mechanical contact under the tension actuators opens. 

\begin{figure}
\includegraphics[width=\columnwidth]{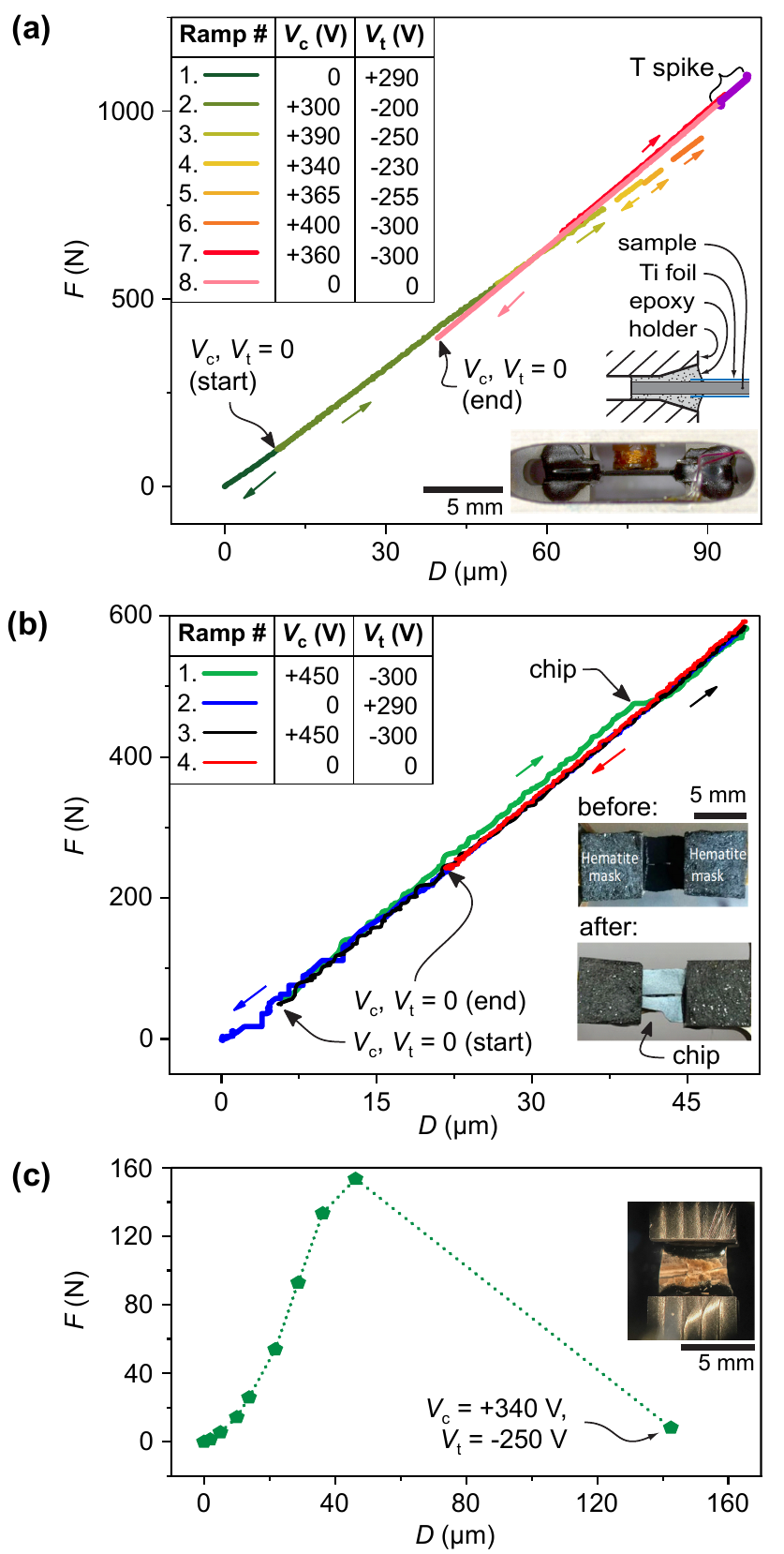}
\caption{Force-displacement traces for three samples. The actuator
voltages were always ramped at temperatures below $10$ K.~\textbf{(a)} A sample of Sr$_2$RuO$_4$ that, within resolution, experienced no damage from application of stress. The table shows the compression and tension actuator voltages $V_\text{c}$ and $V_\text{t}$ at the end of each ramp. The actuators moved further during temperature changes between the voltage ramps, so some of the ramps shown in this figure do not connect with each other. The effect of one temperature spike is shown explicitly. The $F(D)$ traces are all linear, indicating elastic deformation, though between ramps \#6 and 7 a slip appears to have altered the $D$ sensor calibration. Inset: A schematic cross-section through the end of the sample whose data are shown in this panel, and a photograph of the sample from the side. \textbf{(b)} A sample of Sr$_2$RuO$_4$ that fractured partially as force was applied. Inset: Top view of the sample, before and after measurement. \textbf{(c)} A sample of CsFeCl$_3$ that cracked when the load exceeded $\sim$150~N. Inset: Photograph of the sample, top view.}{\label{fig:loading}}
\end{figure}

The above-described cooling procedure was found to consistently place samples under modest compression, in contrast to expectations that the identical thermal contractions of the tension and
compression actuators should keep the load on the sample close to zero.  When the force on the sample is low, the preload force is carried dominantly by
the tension actuators, and we speculate that under this compressive load the tension actuators contract slightly more than the compression actuators upon cooling, causing a
modest force to be applied to the sample.

$F(D)$ data from three samples are illustrated in Fig.~\ref{fig:loading}. Panel (a) shows results from a sample that performed ideally. It was a crystal of Sr$_2$RuO$_4$ with a cross section of $3.1
\times 0.31$~mm$^2$, and to reach the highest possible loads a few extra precautions were taken during mounting. (1) The sample was polished at a 10$^\circ$ angle from the $ab$ planes, so that shear
stress in the ends of the sample would not be aligned with the $ab$ planes, which are cleave planes for Sr$_2$RuO$_4$. (2) Chamfers were cut into the sample holder, as illustrated in
Fig.~\ref{fig:loading}(a). They serve to spread out the stress within the epoxy, reducing peak stress within the epoxy and at the epoxy-sample interface. (3) 10 $\mu$m-thick titanium foils were
affixed to the sample surface, using Stycast\textsuperscript{\textregistered}~1266 epoxy, with the intention of reducing stress concentration at surface defects. The effectiveness of these steps
separately has not been tested. The force applied to this sample was driven over the full range available from the cell without signs of fracture in the sample or epoxy. Multiple ramps of the applied
force are illustrated; within each, $F(D)$ is a straight line. The first ramp was to zero force, and from this the sample was found to have come under a load of 130~N during cooling.  Subsequently,
steadily higher compressive forces were explored. It can be seen that initial and later $F(D)$ curves follow two distinct lines. After the sixth ramp, the temperature spiked to $\sim 12$~K and the
force decreased. We hypothesize that something in the device slipped under the load, for example in the interface between the generator and holder, or in one of the interfaces to the actuator
cartridge: the temperature started to climb at the same time as sharp anomalies in $F$ and $D$ were observed. The $F(D)$ curves may have transitioned to a different slope because the $D$ sensor as
currently designed is not exclusively sensitive to longitudinal displacement, but is also affected by transverse and twisting forces that may have been modified after the slip. Separately, we note
that the sensor readings corresponding to $V_c$, $V_t$ = 0 V differ substantially at the beginning and end of the experiment, due to the hysteresis of the actuators.

Some $F(D)$ traces in Fig.~\ref{fig:loading}(a) do not connect with each other because after the voltage ramps the temperature of the cell was varied for $\mu$SR measurements, and during temperature changes the actuators moved further. An important operational point was revealed: the voltage on the actuators should always be changed at the upper end of the temperature range to be explored. If the actuator voltages are set at a temperature $T_0$, then for all temperatures below $T_0$ the actuators will hold an approximately constant position. When raising temperature above $T_0$ however the actuators may move, and if the voltage is large may do so in abrupt steps that break the sample.

Fig.~\ref{fig:loading}(b) shows $F(D)$ traces for a sample that partially fractured as the applied load was increased. The point where the fracture occurred is clearly visible in the $F(D)$ traces.
Fig.~\ref{fig:loading}(c) shows the $F(D)$ trace for a sample that fractured completely. This sample was CsFeCl$_3$, a gapped quantum magnet \cite{Kurita16}. At an applied force slightly above 150~N, the sample
cracked. $F$ dropped nearly to zero, while $D$ increased sharply. Measurements were continued after the partial fracture of the sample in panel (b), but the fracture of the CsFeCl$_3$ sample was severe enough that the sample needed to be exchanged.

\begin{figure}
\includegraphics[width=\columnwidth]{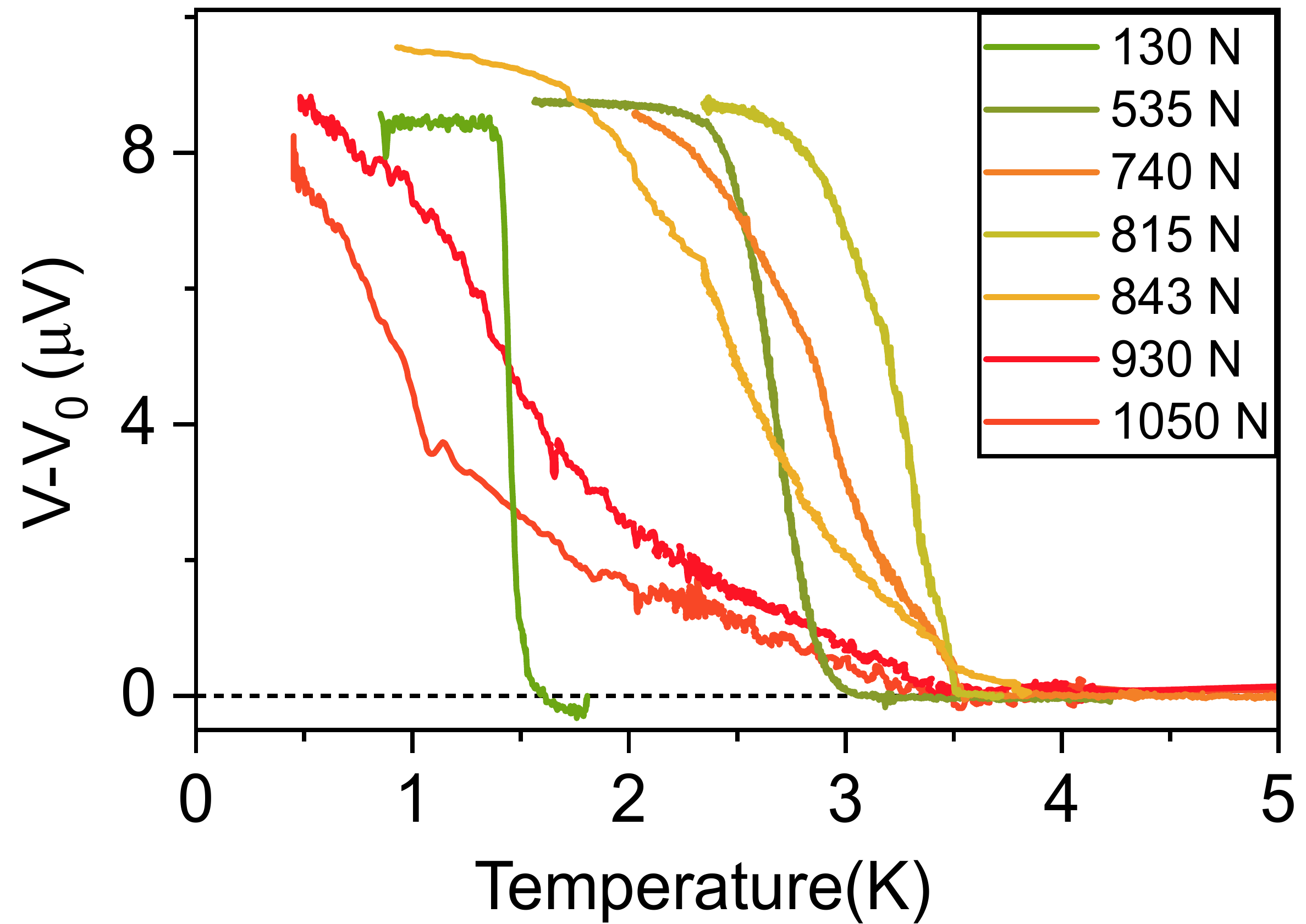}
\caption{ac susceptibility curves of the sample from Fig.~\ref{fig:loading}(a) at various applied forces. The raw Lock-In-amplifier voltage data are shown without normalization, but with subtraction of an offset voltage $V_0$ at each stress. The peak in $T_\text{c}$ was reached at a nominal force of 815~N, which corresponds to a stress of $-0.84 \pm 0.04$~GPa.}
\label{fig:ACS}
\end{figure}

For further discussion of the accuracy of the sensors, we return to Sr$_2$RuO$_4$. At the end of each ramp shown in Fig.~\ref{fig:loading}(a), the magnetic susceptibility of the sample was measured using a pair of coils mounted behind the sample. Results are shown in Fig.~\ref{fig:ACS}: $T_\text{c}$ was found to reach its peak under a load of 815~N. The cross section of the sample was $0.31\times 3.1$~mm$^2$, giving a stress at the maximum $T_\text{c}$ of $-0.84$~GPa. This is in reasonable agreement with the $-0.75 \pm 0.08$~GPa reported in Ref.~\citenum{Barber2019PRB}. $D$ at the peak in
$T_\text{c}$ was 78~$\mu$m. With $k_\text{disp} = 27$~N/$\mu$m, under a load of 815~N the mechanical elements between the displacement sensor and sample compress by 30~$\mu$m, so the
displacement applied to the sample and epoxy is about 48~$\mu$m. The low-temperature Young's modulus of Sr$_2$RuO$_4$ is 160~GPa~\cite{Barber2019PRB}, so at $-0.84$~GPa a strain of $-5.3 \cdot
10^{-3}$ is expected.  $l_\text{exp}$ was 5~mm, so this corresponds to compression of the exposed portion of the sample by 26 $\mu$m. This means that about 22~$\mu$m of the applied displacement went
into the end portions of the sample and the epoxy.

This cell has so far been used both in $\mu$SR and neutron scattering measurements. The Sr$_2$RuO$_4$ sample of Fig.~\ref{fig:loading}(a) had an exposed, pressurized volume of almost 5~mm$^3$ and,
more relevant for $\mu$SR, an area exposed to the beam of 14~mm$^2$. It was driven to a uniaxial stress slightly exceeding 1~GPa. $\mu$SR measurements at the Paul Scherrer Institute showed that at
this stress low-temperature magnetic order was induced in Sr$_2$RuO$_4$~\cite{Grinenko19}.  The sample of Fig.~\ref{fig:loading}(b) had an exposed, pressurized volume of 8 mm$^3$ and exposed area of
20~mm$^2$ (before the partial fracture), and was driven to a maximum uniaxial stress of 430~MPa. $\mu$SR measurements on La$_{1.885}$Ba$_{0.115}$CuO$_4$ under uniaxial stress, using this apparatus,
are reported in Ref.\citenum{Guguchia20}. Finally, CeAuSb$_2$ samples with an exposed, pressurised volume of $\approx 0.6$~mm$^3$ were studied with neutrons at the WISH facility at the ISIS spallation
source, where a uniaxial stress of 440~MPa was reached, and a stress-induced change in the magnetic order was found~\cite{Waite20}. 

\par In summary, we have presented the essential elements for a practical, piezoelectric-driven uniaxial stress cell that can generate stresses exceeding 1~GPa in a sample volume of several mm$^3$.
Cells of the design presented here have accumulated use in multiple beamtimes\cite{Grinenko19, Guguchia20, Waite20}, and we anticipate wide further application.

\begin{acknowledgments}
We thank Jack Barraclough (Razorbill Instruments) and Martin Siegel (TU Dresden) for helpful discussions on design, and the staff of the mechanical workshops at TU Dresden, the Max Planck Institute
for Chemical Physics of Solids, and at the Paul Scherrer Institute for additional discussion and the fabrication of parts. We thank Z. Guguchia and D. Das for technical support during data collection
of Fig.~\ref{fig:loading}(a).  This work has been supported financially by the Deutsche Forschungsgemeinschaft (GR 4667/1, GRK 1621, and SFB 1143), the Max Planck Society, and the Paul Scherrer
Institute. AN acknowledges funding from the European Union’s Horizon 2020 research and innovation program under the Marie Sklodowska-Curie grant agreement No.701647. This work was performed partially
at the Swiss Muon Source (S$\mu$S), PSI, Villigen. CWH has a 31\% ownership stake in Razorbill Instruments, a company that manufactures uniaxial pressure apparatus.  
\end{acknowledgments}

\section*{Appendix}\label{sec:Appendix}
\subsection*{Strain gauge configuration:}
\begin{figure}
\includegraphics[width=\columnwidth]{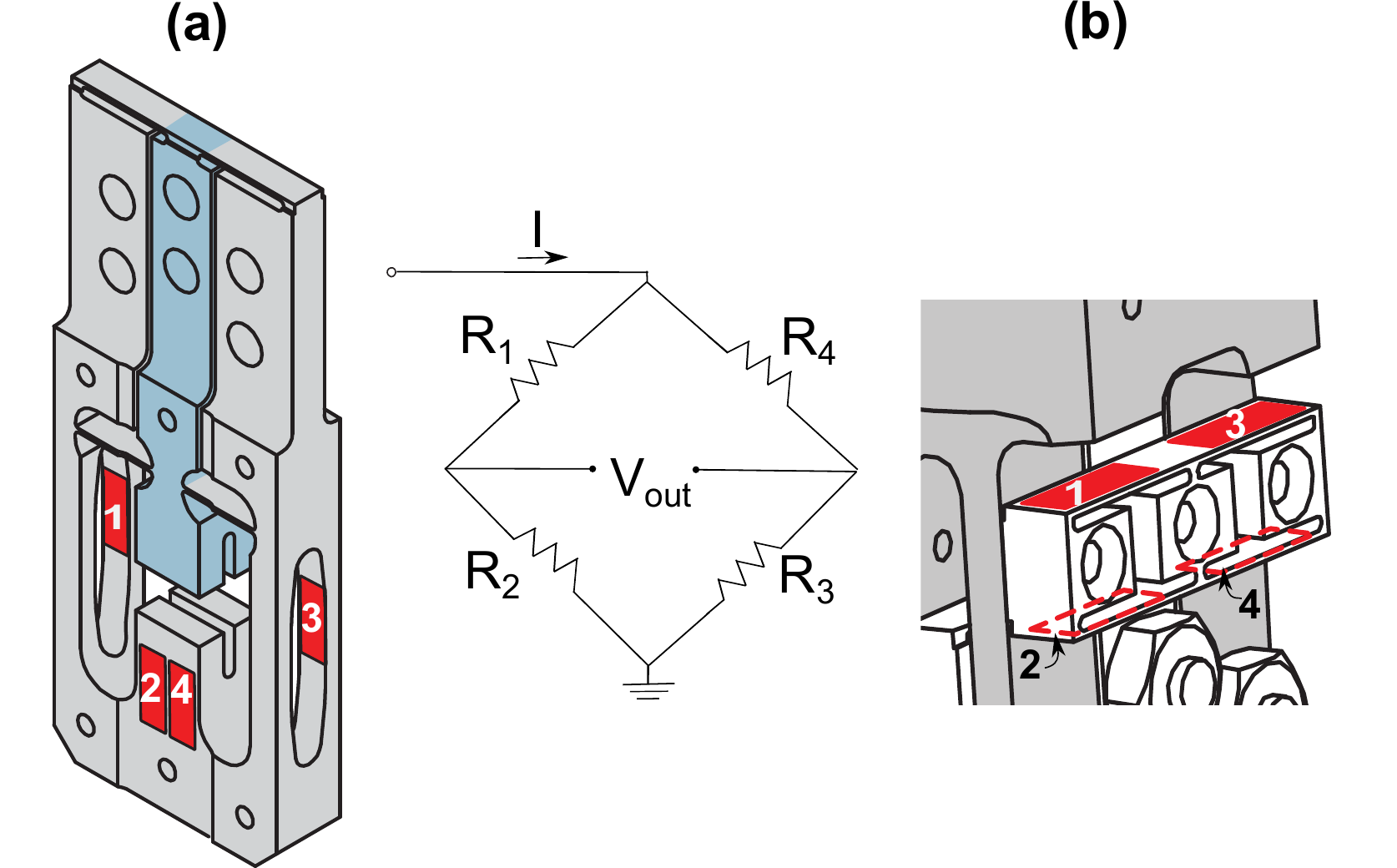}
\caption{Configuration of the sensors. In this cell we have two Wheatstone bridge circuits, \textbf{(a)} force sensor network present in the sample holder and \textbf{(b)} displacement sensor network present in the generator. Each of them are made up of four strain gauges.}
\label{fig:gauges}
\end{figure} 
In this cell, the force and displacement sensors each consist of four strain gauges, connected in a Wheatstone bridge configuration. Fig.~\ref{fig:gauges} shows schematics of the two sensors, including electrical connections. For both sensors, two of the strain gauges ($R_1$ and $R_3$) come under compressive strain when the sample is compressed, and the other two ($R_2$ and $R_4$) under tensile strain. The change in resistance of a single strain gauge is $\Delta R = GR\varepsilon$, where $G$ is the gauge factor, $R$ is the gauge resistance, and $\varepsilon$ is the strain applied to the gauge. For strain gauges connected into a bridge as shown in Fig. \ref{fig:gauges}, this gives

$V_\text{out} = \frac{1}{2} IGR(\varepsilon_1 - \varepsilon_2)$,

where $I$ is the applied current, $\varepsilon_1$ is the strain at gauges 1 and 2, and $\varepsilon_2$ is the strain at gauges 3 and 4. 
\begin{figure}
\includegraphics[width=\columnwidth]{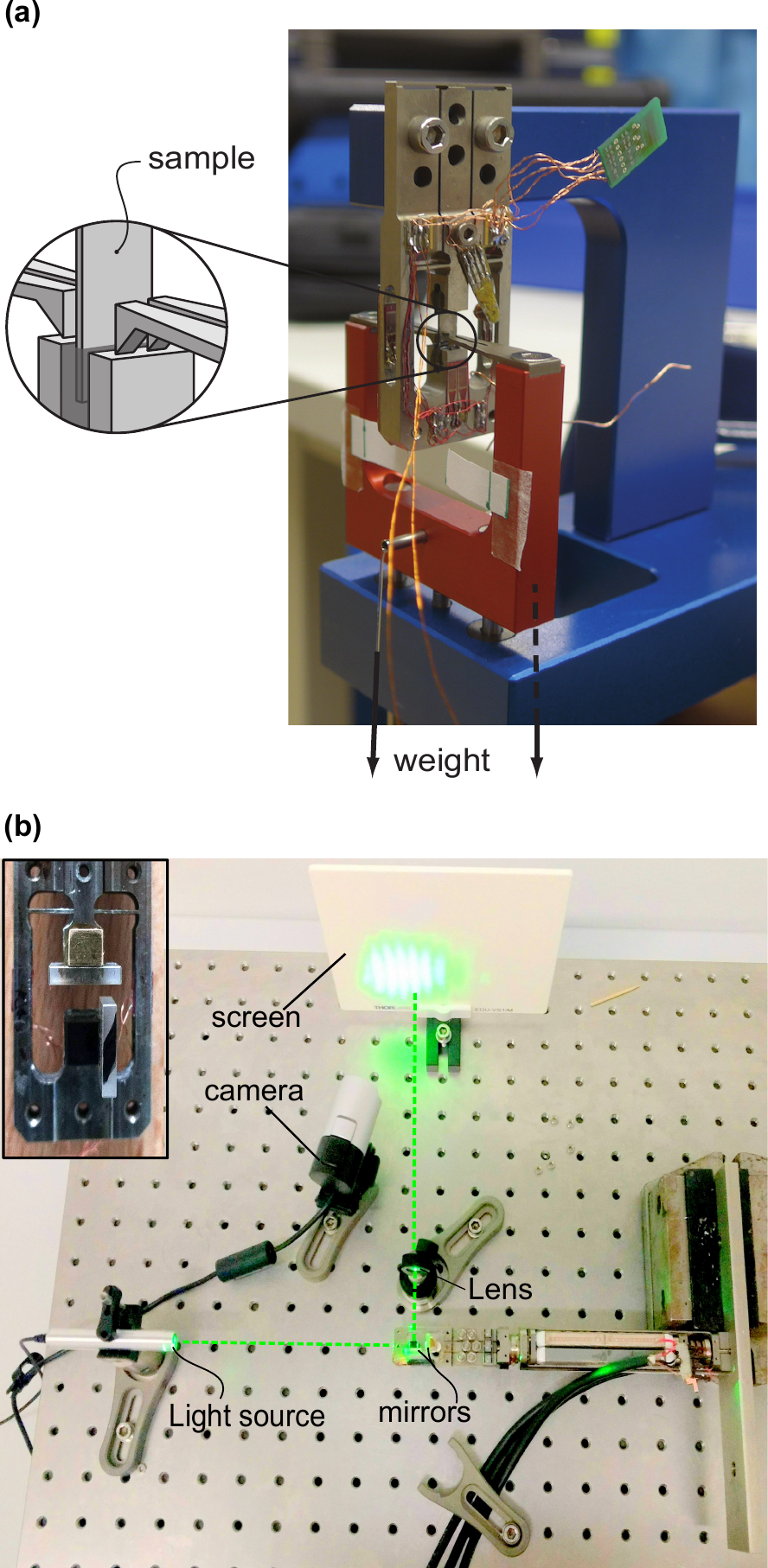}
\caption{Calibration of the $F$ and $D$ sensors. \textbf{(a)} Calibration of the force-sensing strain gauges at room temperature. A known weight is applied to the sample holder, at a location that closely simulates the load from the sample. \textbf{(b)}  Calibration of the displacement sensor using an optical interferometer. Inset: A photograph of the empty sample holder used for calibration. }
\label{fig:calibration}
\end{figure} 
\subsection*{Calibration:}
\par We show in Fig.~\ref{fig:calibration} photographs of the setups we employed for room-temperature calibration of the force and displacement sensors. The photograph in Fig.~\ref{fig:calibration}(a) was taken during  calibration of the force sensors in the holder for Sr$_2$RuO$_4$, whose data are presented in Figs.~\ref{fig:loading}(a) and \ref{fig:ACS}. The setup for the force sensor calibration is designed to apply force in a way that it doesn't pressurize the sample itself. Instead force is applied at points which are expected to give a similar stress profile to the sample. This setup offers the flexibility to perform the force calibration before or after the beamtime. A maximum weight of 1 kg was hung during the calibration. \par For the displacement sensor calibration, we have used a Michelson interferometer (Fig.~\ref{fig:calibration}(b)). A diode pump solid state laser was used as the source ($\lambda$ = 532 nm). A beam splitter and two mirrors were attached on the sample holder, with one of them attached to the movable portion of the holder. The mirrors are shown at the inset of Fig.~\ref{fig:calibration}(b). The interference pattern incident on the screen is recorded by a camera. Up to +25 V were applied to the compression actuators. The optical path difference as well as the phase difference between the reconciling waves changed with the change of applied voltage value,  which was visible as a movement of the sinusoidal pattern on the screen.

\section*{Data Availability Statement}
The data that support the findings of this study are openly available from the Max Planck Digital Library, through the internet link [to be determined].


\begin{thebibliography}{99}
\bibitem{PI} Stack multilayer piezo-actuators (PI Ceramic GmbH, PICMPA\textsuperscript{\textregistered}-887.91) were used, made of a lead zirconate titanate (PZT) piezoelectric material. This actuator
has dimensions 7$\times$7$\times$36~mm.  Two of them were epoxied along the length, to get a 72 mm long stack. 8 actuators were used (in total) per piezo cartridge set.
\end{thebibliography}
\end{document}